\documentclass[sigconf,natbib=true]{acmart}
\AtBeginDocument{%
  }

\makeatletter
\renewcommand{\@copyrightpermission}{}
\makeatother
\setcopyright{acmlicensed}
\copyrightyear{2026}
\acmYear{2026}
\acmDOI{XXXXXXX}
\acmConference[SIGIR '26]{SIGIR}{July 20--24,
  2026}{Melbourne | Naarm, Australia}

\newcommand{\noisename}{hallucinated cues\xspace}

\definecolor{lightgreen}{rgb}{0.0,0.8,0.3}

\newcommand{\zc}[1]{\textcolor{black}{#1}}
\newcommand{\ric}[1]{\textcolor{black}{#1}}

\newcommand{\myheader}[1]{\vspace{0.4em}\noindent\textbf{#1.}}

\newcommand{\ignore}[1]{}

\usepackage[linesnumbered, ruled, vlined]{algorithm2e}
\usepackage{graphicx}
\usepackage{subcaption} 
\usepackage{caption}
\begin{document}

\title{Eliminating Hallucination in Diffusion-Augmented Interactive Text-to-Image Retrieval}

\author{Zhuocheng Zhang}
\affiliation{%
  \institution{Hunan University}
  \city{Changsha}
  \state{Hunan}
  \country{China}
}
\email{zhuocheng@hnu.edu.cn}

\author{Kangheng Liang}
\affiliation{%
  \institution{University of Glasgow}
  \city{Glasgow}
  \country{United Kingdom}
}
\email{2743944l@student.gla.ac.uk}

\author{Guanxuan Li}
\affiliation{%
  \institution{Hunan University}
  \city{Changsha}
  \state{Hunan}
  \country{China}
}
\email{1005890623lgx@gmail.com}

\author{Paul Henderson}
\affiliation{%
  \institution{University of Glasgow}
  \city{Glasgow}
  \country{United Kingdom}
}
\email{Paul.Henderson@glasgow.ac.uk}

\author{Richard Mccreadie}
\affiliation{%
  \institution{University of Glasgow}
  \city{Glasgow}
  \country{United Kingdom}
}
\email{Richard.Mccreadie@glasgow.ac.uk}

\author{Zijun Long$^{\ast}$}
\affiliation{%
  \institution{Hunan University}
  \city{Changsha}
  \state{Hunan}
  \country{China}
}
\thanks{*Corresponding author}
\email{longzijun@hnu.edu.cn}

\renewcommand{\shortauthors}{Trovato et al.}

\begin{abstract}
\looseness -1 Diffusion-Augmented Interactive Text-to-Image Retrieval (DAI-TIR) is a promising paradigm that improves retrieval performance by generating query images via diffusion models and using them as additional ``views” of the user’s intent. However, these generative views can be incorrect because diffusion generation may introduce hallucinated visual cues that conflict with the original query text. Indeed, we empirically demonstrate that these hallucinated cues can substantially degrade DAI-TIR performance. To address this, we propose Diffusion-aware Multi-view Contrastive Learning (DMCL), a hallucination-robust training framework that casts DAI-TIR as joint optimization over representations of query intent and the target image. DMCL introduces semantic-consistency and diffusion-aware contrastive objectives to align textual and diffusion-generated query views while suppressing hallucinated query signals. This yields an encoder that acts as a semantic filter, effectively mapping hallucinated cues into a null space, improving robustness to spurious cues and better representing the user’s intent. Attention visualization and geometric embedding-space analyses corroborate this filtering behavior. Across five standard benchmarks, DMCL delivers consistent improvements in multi-round Hits@10, reaching as high as 7.37\% over prior fine-tuned and zero-shot baselines, which indicates it is a general and robust training framework for DAI-TIR.

\end{abstract}

\maketitle

\section{Introduction}

\begin{figure}[ht!] 
    \centering
    \setlength{\abovecaptionskip}{3pt}
    \setlength{\belowcaptionskip}{-12pt}
    \includegraphics[width=0.9\linewidth]
    {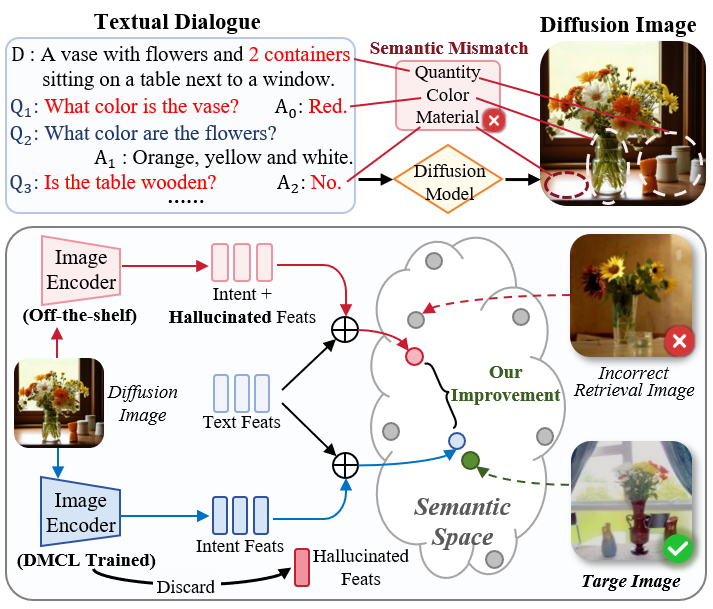}
    \caption{Semantic mismatch between diffusion-generated query images and the textual query due to hallucination.}
    \label{fig:figure1}
\end{figure}
%
Interactive Text-to-Image Retrieval (I-TIR) enables users to iteratively clarify their visual intent through multi-round dialogues with an intelligent agent, allowing the system to progressively align its query representation with the desired target image in real-world scenarios such as fashion recommendation and person search~\cite{levy2023chatting}. Building on recent advances in diffusion-based generation, Diffusion-Augmented Interactive Text-to-Image Retrieval (DAI-TIR) ~\cite{long2025diffusion} uses diffusion models to generate images based on the user's textual query as query-side visual proxies, helping to bridge the gap between users’ linguistic descriptions and their underlying visual intent.

\looseness -1 However, diffusion generated query images can introduce \noisename beyond the user’s intent (Figure \ref{fig:figure1}). Diffusion sampling starts from random noise and produces an image through a denoising trajectory conditioned on the text, so the same prompt can yield multiple visually plausible realizations. When the prompt (the user’s textual query) is underspecified, as is typically the case, the model falls back on learned priors to fill in missing details~\cite{huang2024visual}. This process often fabricates objects, attributes, or backgrounds not mentioned in the textual query, which are referred to as hallucinations. In the context of DAI-TIR, these extra details are not merely harmless variation. They can create conflicting cues with the user’s intended constraints and also shift the generated image and its embedding away from the distribution of real, query relevant images, which together can degrade retrieval.

Despite this, existing DAI-TIR frameworks~\cite{long2025diffusion} typically treat diffusion generated images as ordinary query inputs and apply off-the-shelf retrieval encoders that were pre-trained or fine-tuned on natural image--text pairs. These encoders are optimized for relatively clean and semantically consistent pairs, rather than for diffusion outputs that exhibit distribution shift and occasional hallucinations. As a consequence, they neither detect nor correct cross-view conflicts, and they can be distracted by hallucinated visual details. We argue therefore that it is essential for DAI-TIR solutions to train the encoder to identify these \noisename and suppress them. Furthermore, DAI-TIR systems should ideally enforce semantic consistency across all views of the query intention and the target image, so that they occupy a mutually aligned region of the embedding space. Without such joint consistency, diffusion based views can dilute the original intent expressed in the text, leading to unstable matching behavior and limiting the benefits of diffusion augmentation.

To address these limitations, we propose Diffusion-aware Multi-View Contrastive Learning (DMCL), a hallucination-robust training framework for DAI-TIR. DMCL casts DAI-TIR as a joint optimization problem that enforces semantic consistency across multiple representations of the query. It introduces multi-view contrastive objectives~\cite{zhang2022contrastive, goel2022cyclip} that explicitly improve the alignment between different query views and the target image. By tying these views together during training, DMCL suppresses \noisename introduced by the diffusion model and improves retrieval performance.

For each dialogue round, DMCL builds three complementary query representations (Figure~\ref{fig:figure2}): a text query that specifies the user’s query intention, including semantics and constraints; a diffusion-based query that acts as a visual proxy of the user’s intention; and a fused query that combines both signals.
Building upon these representations, DMCL enhances DAI-TIR with two complementary training objectives: Multi-View Query–Target Alignment objective and Text–Diffusion Consistency objective.
The multi-view query–target alignment objective trains the retriever to align all query views with the same target image, encouraging it to emphasize query cues shared across views while filtering out hallucinated or inconsistent details introduced by diffusion generation.
The consistency objective explicitly enhances agreement between the text query and the diffusion query making the model less sensitive to text–diffusion mismatches and generative hallucination.
Together, both objectives shape a shared embedding space that works reliably for the DAI-TIR task even under multi-round user intent shifts.



Our main contributions are summarized as follows:
\begin{itemize}
\item We show and quantify that diffusion-generated query images can introduce hallucinated content into query representations, which can substantially degrade retrieval performance.
\item We propose \emph{Diffusion-aware Multi-View Contrastive Learning} (DMCL), a hallucination-robust training framework that encourages semantic consistency across all query views, enabling the encoder to suppress diffusion-induced \noisename during retrieval.
\item We conduct extensive experiments on five I-TIR dialogue benchmarks under both in-distribution and out-of-distribution evaluation settings. DMCL consistently improves multi-round Hits@10 over existing methods, demonstrating both effectiveness and generalizability.
\item We provide empirical evidence that DMCL acts as a semantic filter by reshaping the embedding space, improving cross-modal alignment among query views while suppressing conflicting signals.
\item We release a large-scale DAI-TIR dataset to facilitate future research.
\end{itemize}

\section{Related Work}

\looseness -1 Interactive Text-to-Image Retrieval (I-TIR) extends standard retrieval paradigms by allowing users to refine their search intent through multi-round interactions. Early approaches relied on relevance feedback mechanisms, where users explicitly marked retrieved images as relevant or irrelevant to update the ranking model ~\cite{rui1998relevance, zhou2003relevance}. 
Following the shift toward generative paradigms~\cite{nichol2021glide, saharia2022photorealistic, ramesh2022hierarchical}, recent I-TIR research explored LLM-based textual augmentation. Methods such as ChatIR~\cite{levy2023chatting} and PlugIR~\cite{lee2024interactive} rewrite multi-round dialogue contexts into retriever-friendly captions, thereby improving linguistic clarity and reducing conversational ambiguity. Yet, because they remain strictly within the textual modality, they cannot explicitly reconstruct missing visual attributes (e.g., color, texture) that are often crucial for fine-grained matching to the image corpus.

\looseness -1 Motivated by this limitation, recent work has turned to Diffusion-
Augmented Interactive Text-to-Image Retrieval (DAI-TIR)~\cite{long2025diffusion}, enabled by rapid advances in diffusion probabilistic models~\cite{ho2020denoising, rombach2022high}. Rather than only refining text, DAI-TIR synthesizes one or multiple visual proxies of the user’s intent~\cite{long2025diffusion}, based on the premise that generated images can provide a concrete visual anchor that complements—and sometimes goes beyond—verbal descriptions. Indeed, Long et al.~\cite{long2025diffusion} show that augmenting text queries with synthetic images improves retrieval performance even in zero-shot settings. However, generative models introduce a key challenge: when user intent is underspecified, diffusion models fall back on learned priors to hallucinate missing details—effectively making educated guesses to complete the image~\cite{qin2024evaluating}. Although these synthesized details provide visual plausibility, standard image encoders will then often treat these details as hard constraints rather than tentative hypotheses ~\cite{huang2023t2i,lim2025evaluating}. We show that this can mislead retrieval, as the retriever becomes distracted by \noisename (e.g., background elements or object style) that were never specified by the user. Our work further addresses this limitation not by discarding generation, but by tuning the encoder such that it can distinguish between core semantic intent and these conflicting generative details.

\section{Are Diffusion-added Hallucinated Cues a Problem?}

Before presenting our approach for mitigating \noisename introduced by diffusion generated query images, we first formalize the task and quantify the prevalence of such problematic cues.

\subsection{Task Definitions}
\label{subsec:da-itir}
\looseness -1 \noindent \textbf{Interactive Text-to-Image Retrieval (I-TIR)}: I-TIR is a multi-round task that aims to identify a target image $I^* \in \mathcal{I}$ through an iterative dialogue between a user and a retrieval system. The interaction starts with an initial description $D_0$ that provides a coarse specification of the target image. At each round $n$, the system asks a clarifying question $Q_n$, the user replies with an answer $A_n$, and the dialogue context up to round $n$ is denoted as:\vspace{-1mm}
\begin{equation}
C_n = \{D_0,Q_1,A_1,...,Q_n,A_n\}
\label{C_t}
\end{equation}
This dialogue context undergoes processing steps, such as concatenating all text elements, to form a unified text search query $T_n$. 



\vspace{2mm}\noindent \textbf{Diffusion-Augmented Interactive Text-to-Image Retrieval (DAI-TIR)}:
In DAI-TIR, the dialogue context $C_t$ is reformulated into a prompt $S_n$ and fed to a pre-trained text-to-image diffusion model $\mathcal{G}$ to generate a set of proxy query images $\{I_n^{(k)}\}_{k=1}^{K}$, where $I_n^{(k)}=\mathcal{G}(S_t;\epsilon_k)$ corresponds to an independent diffusion sample (e.g., different noise seeds). These generated images serve as visual proxies of the user’s current intent and are incorporated into retrieval. The underlying assumption is that multiple diffusion-generated proxies can provide richer and more informative representations of the user’s information need than text alone, thereby improving retrieval performance.

To obtain a unified query representation, a fusion function $\mathcal{F}(\cdot,\cdot)$ combines the text embedding $T_n$ and the set of proxy image embeddings $\{I_n^{(k)}\}_{k=1}^{K}$ to produce a diffusion-augmented query:
\begin{equation}
    F_n = \mathcal{F}\!\left(T_n, \{I_n^{(k)}\}_{k=1}^{K}\right).
\end{equation}
\looseness -1 The retriever then scores each candidate image $I \in \mathcal{I}$ by similarity $s(I, F_n)$ and ranks them accordingly. This procedure repeats across dialogue rounds until the target image is retrieved or a preset interaction limit is reached. Formally, the final retrieval result is given by:
\begin{equation}
    I^* = \arg\max_{I \in \mathcal{I}} s(I, F_n).
\end{equation}

The performance of the retrieval system can be evaluated based on the retrieval rank of the target image.

\begin{figure}[t]
    \centering
    
    \begin{subfigure}[b]{0.4\linewidth} 
        \centering
        \includegraphics[width=\linewidth]{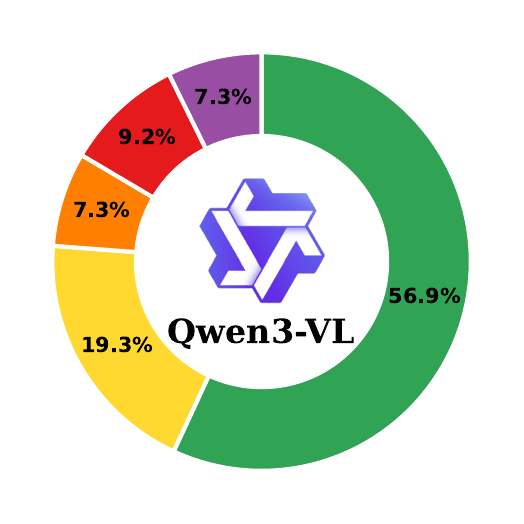}
    \end{subfigure}
    \hspace{-3mm}
    \begin{subfigure}[b]{0.4\linewidth}
        \centering
        \includegraphics[width=\linewidth]{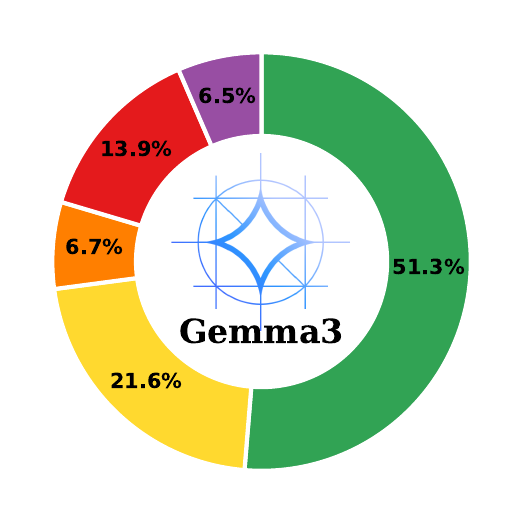}
    \end{subfigure}
    
    \vspace{-3mm}
    
    \begin{subfigure}[b]{0.7\linewidth}
        \centering
        \includegraphics[width=\linewidth]{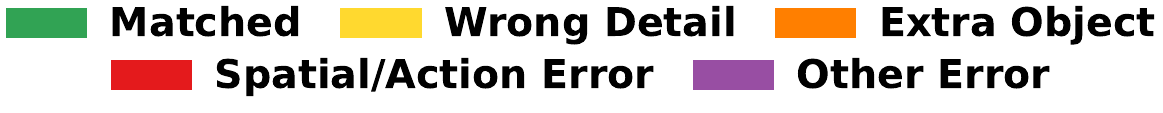}
    \end{subfigure}
    \vspace{-3mm}
    
    \caption{Text consistency and hallucination-type distribution of diffusion-generated images, evaluated by Qwen3-VL and Gemma3.}
    \label{fig:genImg_noise}
    \vspace{-3mm}
\end{figure}

\subsection{Hallucinated Cues in Diffusion-Generated Images}
Motivated by the concern that DAI-TIR may be undermined by spurious or conflicting visual details in generated proxies, we quantify how often diffusion generated query images introduce \noisename or more serious conflicts with the initial query intent within the DAI-TIR setting. We develop an automatic evaluation protocol that uses large multimodal models as judges. Specifically, we employ Qwen3 VL~\cite{yang2025qwen3} and Gemma3~\cite{team2025gemma} to evaluate whether each generated image is consistent with the corresponding dialogue context. We apply a Chain of Thought prompting strategy~\cite{wei2022chain} to encourage grounded, step by step reasoning during judgment. Each case is then assigned to one of the following categories:
\begin{itemize}
    \item \textbf{Matched:} The generated image is semantically consistent with the dialogue text.
    \item \textbf{Wrong Detail:} The generated image contradicts the text in visual attributes (e.g., color, count, shape).
    \item \textbf{Extra Object:} The generated image contains prominent distracting objects not mentioned in the text.
    \item \textbf{Spatial/Action Error:} The image conflicts with the text in spatial relations or actions.
    \item \textbf{Other Error:} The image violates other constraints (e.g., physical plausibility, scene setting).
\end{itemize}

Figure~\ref{fig:genImg_noise} reports the proportion of generated images matched to each of the above classes. Although these AI-judges will be error-prone (and as such we should not read too much into the exact ratios), we observe broadly consistent behavior from both judges, i.e. around 40\% of diffusion generated images contain inconsistencies with the ground truth dialogue text, demonstrating that this is likely an important problem. These inconsistencies can introduce misleading visual cues that contaminate the query representation, weaken the retriever’s ability to discriminate relevant candidates, and ultimately degrade retrieval performance. This observation motivates equipping the encoder with mechanisms to suppress influence of \noisename during retrieval.

\section{Reducing Hallucinated Cues}
\label{sec:methods}
\subsection{How to suppress diffusion \noisename?}
\label{subsec:why_filter}
Diffusion generated query images can be valuable because they offer an explicit visual view of the user’s intent. However, they are also can be unreliable: when the query is underspecified, diffusion models fall back on learned priors to fill in missing details~\cite{chefer2023attend}, often hallucinating objects, attributes, or backgrounds that the user never specified. Such spurious content can conflict with the intended constraints and introduce a distribution shift away from real query relevant images, which can degrade retrieval performance~\cite{huang2024visual, yarom2023you}. This motivates a training objective that makes the encoder sensitive to intent-relevant semantics while ignoring to diffusion hallucinated cues.

\begin{figure*} 
    \centering
    \includegraphics[width=.95\textwidth]{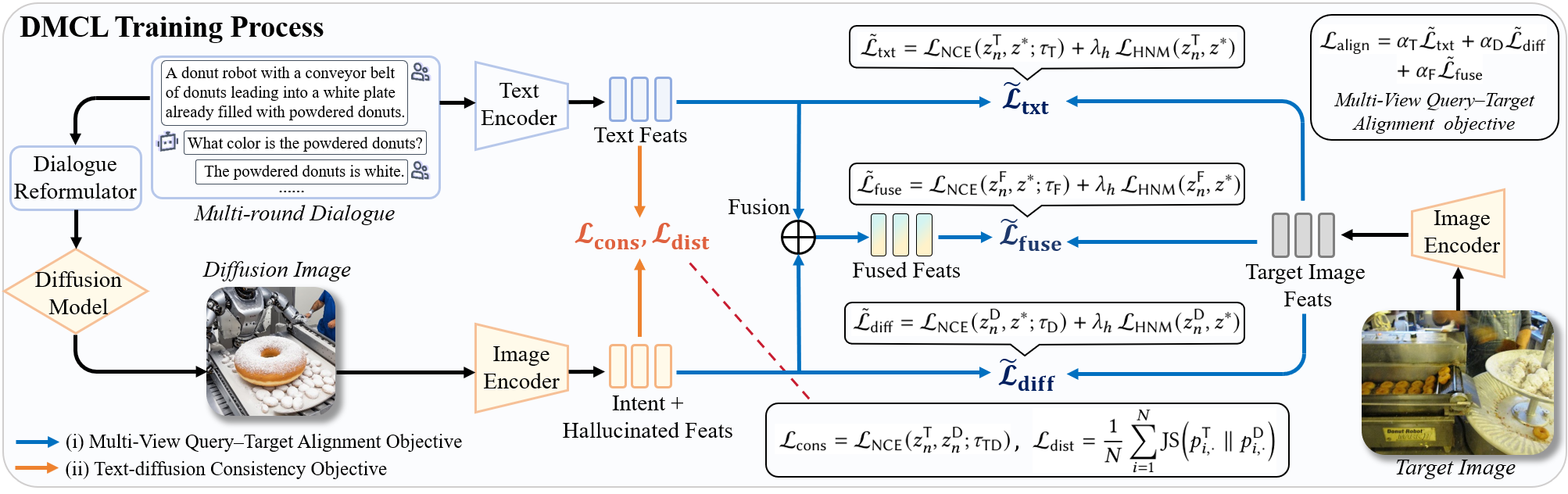}
    \vspace{-3mm}
    \caption{DMCL Training Framework Overview}
    \vspace{-3mm}
    \label{fig:figure2} 
\end{figure*}

Intuitively, intent-consistent details should be supported by the dialogue text and remain stable across interactive rounds; hallucinated details should vary across rounds and lack consistent textual support. In a well trained embedding space, different semantic factors often combine approximately linearly, so multiple sources of evidence can be modeled as approximately additive components in the final representation~\cite{mikolov2013efficient, radford2015unsupervised, radford2021learning}. Here, we note that this is an assumption about late layer feature space, where vector addition is frequently a reasonable approximation for composing attributes in well optimized contrastive encoders.

Concretely, let $x_{\text{gen}}$ denote a late layer feature (or the pre-projection embedding) extracted from the diffusion proxy image. We assume an additive decomposition in feature space:
\begin{equation}
x_{\text{gen}} = s + h,
\end{equation}
where $s$ captures dialogue-consistent semantics (intent-relevant content that is supported by the text and stable across diffusion samples), and $h$ captures diffusion-induced variation such as hallucinated attributes, incidental backgrounds, or stylistic details (which tend to vary across random seeds and are not reliably supported by the text). Since the text view is a cleaner expression of intent, we treat it as primarily encoding the same stable component:
\begin{equation}
x_{\text{text}} \approx s.
\end{equation}
Our goal is to learn an encoder $f_\theta$ that maps an input view to an embedding used for retrieval, preserving $s$ while suppressing $h$. A natural mechanism is an alignment objective that pulls embeddings of semantically consistent views together while separating mismatched pairs.

\vspace{2mm}\noindent \textbf{Why alignment prefers the stable component}:
Consider an alignment loss based on cosine similarity between a diffusion view and its text anchor:
\begin{equation}
\mathcal{L}_{\text{cos}}(x_{\text{gen}}, x_{\text{text}})
= - \cos\!\big(f_\theta(x_{\text{gen}}), f_\theta(x_{\text{text}})\big),
\end{equation}
where the negative sign indicates that minimizing the loss maximizes cosine similarity. To expose the effect of diffusion variation, we use a local linear approximation of the encoder around the relevant region of feature space, $f_\theta(x) \approx W x$, where $W$ is the Jacobian like linear map. Define the (unnormalized) embeddings
\begin{equation}
q_{\text{gen}} = W(s+h), \qquad q_{\text{text}} = Ws,
\end{equation}
noting that cosine similarity implicitly normalizes by the vector norms.
The numerator of the cosine similarity is
\begin{equation}
q_{\text{gen}}^\top q_{\text{text}}
= (Ws + Wn)^\top (Ws)
= \|Ws\|^2 + (Wn)^\top (Ws).
\end{equation}
If diffusion induced variation is not systematically correlated with the text for a fixed intent, then
$\mathbb{E}[(Wn)^\top (Ws)] = 0$, where the expectation $\mathbb{E}[\cdot]$ is taken over diffusion randomness (e.g., noise seeds) conditioned on the same underlying intent $s$. Meanwhile, the denominator includes the noise energy:
\begin{equation}
\|q_{\text{gen}}\|^2 = \|Ws\|^2 + \|Wn\|^2 + 2(Wn)^\top (Ws).
\end{equation}
Therefore, in expectation the cosine satisfies
\begin{equation}
\scalebox{0.95}{$
\mathbb{E}\!\left[\cos(q_{\text{gen}}, q_{\text{text}})\right]
\approx
\frac{\|Ws\|^2}{\sqrt{(\|Ws\|^2 + \mathbb{E}\|Wn\|^2)\,\|Ws\|^2}} 
=
\frac{1}{\sqrt{1 + \mathbb{E}\|Wn\|^2 / \|Ws\|^2}}
$}
\label{eq:cos_expect}
\end{equation}


\vspace{2mm}\noindent Maximizing alignment therefore encourages $\mathbb{E}\|Wn\|^2$ to be small relative to $\|Ws\|^2$. In other words, representation directions dominated by $h$ reduces expected agreement with the text anchor, so training pushes such directions toward a near null response. This matches the desired semantic filtering behavior.

\vspace{2mm}\noindent \textbf{From alignment to contrastive objectives}:
In retrieval, the encoder must not only align matched items but also separate them from confusable candidates. To satisfy both requirements, the encoder is encouraged to represent what is consistently predictive of the correct match across views, and to ignore what is unstable and non-predictive~\cite{oord2018representation, karpukhin2020dense, qu2025intent}. In practice, this biases the representation toward the intention component $s$ and away from the hallucination component $h$.
Let $I^+$ denote the target image and $\mathcal{B}$ denote the set of images in a minibatch. For a query view $v$ with embedding $q^{(v)}$ and image embedding $k(I)$, a standard contrastive retrieval objective~\cite{ chen2025dynamic, dong2024unsupervised, xu2024cmclrec, zhu2023adamcl} is
\begin{equation}
\mathcal{L}_{\text{ret}}^{(v)}
= - \log
\frac{\exp\!\left(\mathrm{sim}(q^{(v)}, k(I^+))/\tau\right)}
{\sum_{I \in \mathcal{B}} \exp\!\left(\mathrm{sim}(q^{(v)}, k(I))/\tau\right)}.
\label{eq:single_contrastiveloss}
\end{equation}
We can apply this loss to multiple query views, for example a text query, a diffusion query, and a fused query, and sum them
\begin{equation}
\mathcal{L}_{\text{multi}}
= \sum_{v \in \mathcal{V}} \mathcal{L}_{\text{ret}}^{(v)}.
\end{equation}
This jointly rewards what is consistently predictive of the target across views, and it penalizes features like $h$ that fluctuate, because they do not provide reliable separation from negatives~\cite{zhao2023keyword, zhang2022image, wang2022improving}.

Finally, to reduce severe mismatch between text and diffusion views without forcing them to collapse, we add a cross view consistency term, written in the same contrastive form
\begin{equation}
\mathcal{L}_{\text{cons}}
= - \log
\frac{\exp\!\left(\mathrm{sim}(q^{(\text{text})}, q^{(\text{gen})})/\tau\right)}
{\sum_{q \in \mathcal{Q}} \exp\!\left(\mathrm{sim}(q^{(\text{text})}, q)/\tau\right)}.
\end{equation}
Overall, minimizing
\begin{equation}
\mathcal{L} = \mathcal{L}_{\text{multi}} + \lambda\, \mathcal{L}_{\text{cons}}
\end{equation}
encourages a representation space where intent related semantics are aligned across views and to the target, while diffusion specific \noisename is suppressed because it is unstable and not consistently rewarded by the contrastive signal.

From this perspective, the desired semantic filtering behavior is directly imposed by the designed contrastive constraints across views.
By treating text, diffusion proxies, fused queries, and target images as multiple observations of the same underlying intent, and by enforcing their agreement while discriminating against confusable alternatives, we can guide the encoder to preserve the consistent semantics and suppress diffusion specific \noisename. This reasoning directly motivates multi-view contrastive objectives that coordinate the query views with the target and also encourage agreement among the query views themselves.

\subsection{Proposed Approach: DMCL}
\label{subsec:dmcl}
\looseness -1 DMCL is a diffusion-hallucination-aware contrastive training framework for DAI-TIR that learns a shared embedding space where intent-consistent semantics are reinforced across views, while diffusion specific \noisename are suppressed. As shown in Figure~\ref{fig:figure2}, during training, we first construct three query-side views at each dialogue round, namely text, diffusion proxy, and a fused view (Section~\ref{subsubsec:dmcl_repr}). We then optimize two complementary objectives: (i) a multi-view query--target alignment objective \zc{(indicated by blue arrows in Figure~\ref{fig:figure2})} that aligns each query view to the ground-truth target image using a generalized symmetric contrastive objective (Section~\ref{subsubsec:dmcl_nce}) together with hard-negative mining (Section~\ref{subsubsec:dmcl_hnm}), formalized in Section~\ref{subsubsec:dmcl_align}; and (ii) a text--diffusion consistency objective \zc{(indicated by orange arrows in Figure~\ref{fig:figure2})} that explicitly couples the textual query with its generated proxy to reduce cross-view drift (Section~\ref{subsubsec:dmcl_cons}). Finally, we combine both objectives into a overall training objective in Section~\ref{subsubsec:dmcl_total}. At inference, we embed all query views and retrieve by nearest-neighbor similarity in the learned space.

\subsubsection{Multi-View Construction}
\label{subsubsec:dmcl_repr}

For the sample pair $ i $ in a dataset or a mini batch, at dialogue round $n$, we denote the dialogue-conditioned textual query (e.g., reformulated intent and explicit constraints) as $T_{n,i}$, and the corresponding diffusion-generated visual proxy as $I_{n,i}$. To explicitly model complementary intent signals, we decouple the query into three query-side views and embed them into a shared $d$-dimensional representation space: a textual view, a diffusion visual view, and a fused synergistic view. 

Concretely, for each view $v\in\{\textsf{T},\textsf{D},\textsf{F}\}$, we use a view-specific encoder $\Phi_v(\cdot)$, which is composed of a backbone followed by a lightweight projection head (e.g., linear layer or MLP) to map features into the common embedding space. A fusion function $\mathcal{F}_{\textsf{F}}(\cdot,\cdot)$ combines the textual view $\textsf{T}$ and the diffusion view $\textsf{D}$ to produce a fused synergistic view $\textsf{F}$. For a mini-batch of size $N$, the three query embeddings at round $n$ are computed as:
\begin{equation}
    z_{n,i}^{\textsf{T}} = \Phi_{\textsf{T}}(T_{n,i}),\quad
    z_{n,i}^{\textsf{D}} = \Phi_{\textsf{D}}(I_{n,i}),\quad
    z_{n,i}^{\textsf{F}} = \mathcal{F}_{\textsf{F}}(T_{n,i}, I_{n,i}),
\end{equation}
where $z_{n,i}^{\textsf{T}}$, $z_{n,i}^{\textsf{D}}$, and $z_{n,i}^{\textsf{F}}$ correspond to the text, diffusion, and fused views, respectively. 

On the target side, we embed the ground-truth image $I^*$ using an image encoder $\Phi_{\textsf{I}}(\cdot)$ 
\begin{equation}
    z^{*}_i = \Phi_{\textsf{I}}(I^*_i).
\end{equation}

All embeddings are $\ell_2$-normalized and similarities are computed by dot product $s(a,b)=a^\top b$.

\subsubsection{Transition to Generalized Symmetric Contrastive Objective}
\label{subsubsec:dmcl_nce}

For each instance $i$ , the text, diffusion, and fused view, together with the target image, are treated as a multi-positive set that reflects the same underlying intent. The standard InfoNCE loss in Eq.~\ref{eq:single_contrastiveloss} assumes a single positive for each anchor and is typically applied in a one view to one target manner, which makes it difficult to explicitly express one to many positive relations across views. To allow the model to distribute probability mass over multiple valid positives and to avoid asymmetric training effects, we adopt a symmetric InfoNCE formulation~\cite{robinson2020contrastive, chuang2020debiased, chuang2022robust} in Eq.~\ref{eq:snce}:
\begin{equation}
\mathcal{L}_{\textsf{NCE}}(U,V;\tau) =
-\frac{1}{2N}\Bigg(
\sum_{i=1}^{N}\sum_{j=1}^{N}\tilde{q}_{i,j}\log p_{i,j}
+
\sum_{i=1}^{N}\sum_{j=1}^{N}\tilde{q}_{j,i}\log p_{j,i}
\Bigg)
\label{eq:snce}
\end{equation}
where $q_{i,j}\in\{0,1\}$ indicates whether $(u_i,v_j)$ is a positive pair from a multi-view positive set and
\begin{equation}
    \hat{q}_{i,j}=\frac{q_{i,j}}{\sum_{k=1}^N q_{i,k}}\qquad
    \tilde{q}_{i,j}=(1-\epsilon)\hat{q}_{i,j}+\epsilon\cdot\frac{1}{N}
\end{equation}
with $\epsilon\in[0,1)$ denoting label smoothing. Predicted match probabilities are
\begin{equation}
    p_{i,j} = \frac{\exp(s(u_i,v_j)/\tau)}{\sum_{k=1}^{N}\exp(s(u_i,v_k)/\tau)}\qquad
    p_{j,i} = \frac{\exp(s(v_i,u_j)/\tau)}{\sum_{k=1}^{N}\exp(s(v_i,u_k)/\tau)}
\end{equation}

This objective supports a set of positives per anchor across views, and improves robustness via label smoothing and bidirectional optimization. Consequently, it acts as a semantic filter that suppresses view-specific conflicting cues while retaining intent signals that are consistent across views.




\subsubsection{Hard-Negative Mining}
\label{subsubsec:dmcl_hnm}

In practice, diffusion augmentation can introduce spurious attributes that make certain irrelevant target images appear highly similar to a query view, producing a small set of \emph{highly confusable} negatives.
While $\mathcal{L}_{\textsf{NCE}}$ uses all in-batch negatives, its gradients are often dominated by many easy negatives and may under-emphasize these confusable cases.
To directly target the failure mode induced by diffusion hallucinations, we incorporate a hard-negative mining (HNM) objective that prioritizes separating each anchor from its most similar (and thus most misleading) negatives.
By repeatedly pushing away negatives that are close \emph{because of spurious, view-specific cues}, HNM encourages the encoder to rely on signals that are consistent across views.
Consequently, it reinforces DMCL's semantic-filtering effect: retaining intent-consistent signals while suppressing hallucination-driven similarities.

\looseness -1 For an anchor $u_i$, we define positives $\mathcal{P}_i=\{j\mid q_{i,j}=1\}$ and negatives $\mathcal{N}_i=\{j\mid q_{i,j}=0\}$, then mine the top-$K$ most similar negatives:
\begin{equation}
\mathcal{H}_i = \mathrm{TopK}\big(\{\, s(u_i,v_j)\,:\, j\in\mathcal{N}_i \,\}, K\big)
\label{eq:hardset}
\end{equation}
For multi-positive cases, we use the pooled positive similarity
\begin{equation}
    s_i^{+}=\frac{1}{|\mathcal{P}_i|}\sum_{p\in\mathcal{P}_i} s(u_i,v_p)
\end{equation}
We then enforce a margin $m>0$ via a soft ranking loss:
\begin{equation}
\mathcal{L}_{\textsf{HNM}}^{u\rightarrow v}(U,V)
=
\frac{1}{N}\sum_{i=1}^{N}
\frac{1}{|\mathcal{H}_i|}\sum_{j\in\mathcal{H}_i}
\log\Big(1+\exp\big(\tfrac{s(u_i,v_j)-s_i^{+}+m}{\tau_h}\big)\Big)
\label{eq:hnm_uv}
\end{equation}
where $\tau_h$ is the temperature controlling hard-negative separation.
We define $\mathcal{L}_{\textsf{HNM}}^{v\rightarrow u}$ analogously by swapping the roles of $U$ and $V$ and mining hard negatives for anchors in $V$.
The final symmetric HNM term is
\begin{equation}
\mathcal{L}_{\textsf{HNM}}(U,V)
=
\frac{1}{2}\Big(\mathcal{L}_{\textsf{HNM}}^{u\rightarrow v}(U,V)+\mathcal{L}_{\textsf{HNM}}^{v\rightarrow u}(V,U)\Big)
\label{eq:hnm_sym}
\end{equation}

\subsubsection{Multi-View Query--Target Alignment Objective}
\label{subsubsec:dmcl_align}

The multi-view query-target alignment objective (indicated by blue arrows in Figure~\ref{fig:figure2}) explicitly align \emph{each} query view to the same ground-truth target image $I^*$ .
This shared supervision encourages agreement on semantics that are consistent across views and discourages reliance on view-specific spurious cues.
We implement this as a composite objective (contrastive alignment + hard-negative enhancement) for text, diffusion, and their fused representation.

\myheader{Text--Target Alignment}
This objective ensures that the backbone encoder faithfully captures the dialogue-conditioned intent and explicit linguistic constraints in $T_n$:
\begin{equation}
\tilde{\mathcal{L}}_{\textsf{txt}}
=
\mathcal{L}_{\textsf{NCE}}(z_{n}^{\textsf{T}}, z^{*}; \tau_{\textsf{T}})
+\lambda_h\,\mathcal{L}_{\textsf{HNM}}(z_{n}^{\textsf{T}}, z^{*})
\end{equation}

\myheader{Diffusion--Target Alignment}
We treat the diffusion-generated image $I_n$ as a visual proxy for the user's mental imagery and align it to the same target $I^*$:
\begin{equation}
\tilde{\mathcal{L}}_{\textsf{diff}}
=
\mathcal{L}_{\textsf{NCE}}(z_n^{\textsf{D}}, z^{*}; \tau_{\textsf{D}})
+\lambda_h\,\mathcal{L}_{\textsf{HNM}}(z_n^{\textsf{D}}, z^{*})
\end{equation}

Because $I_n$ may contain hallucinated details, enforcing target agreement (and separating hard negatives) pressures the encoder to retain target-relevant semantics while down-weighting hallucination-driven cues that do not consistently explain $I^*$.

\myheader{Fused--Target Alignment}
We optimize the integrated representation $F_n$, dynamically fused from $T_n$ and $I_n$, to emphasize complementary semantics while remaining anchored to the ground-truth:
\begin{equation}
\tilde{\mathcal{L}}_{\textsf{fuse}}
=
\mathcal{L}_{\textsf{NCE}}(z_n^{\textsf{F}}, z^{*}; \tau_{\textsf{F}})
+\lambda_h\,\mathcal{L}_{\textsf{HNM}}(z_n^{\textsf{F}}, z^{*})     
\end{equation}

\myheader{Weighted Multi-View Alignment}
Finally, we combine the three alignment objectives with learnable view weights:
\begin{equation}
    \mathcal{L}_{\textsf{align}}
    =
    \alpha_{\textsf{T}}\tilde{\mathcal{L}}_{\textsf{txt}}
    + \alpha_{\textsf{D}}\tilde{\mathcal{L}}_{\textsf{diff}}
    + \alpha_{\textsf{F}}\tilde{\mathcal{L}}_{\textsf{fuse}},
\end{equation}
where $\alpha_{\textsf{T}},\alpha_{\textsf{D}},\alpha_{\textsf{F}}>0$ are learnable scalars and $\lambda_h$ controls the strength of hard-negative mining.
By adaptively weighting the views, the model can rely more on the most reliable representation for a given query instance (e.g., down-weighting the diffusion view when it is unreliable), further improving robustness.

\subsubsection{Text--Diffusion Consistency Objective}
\label{subsubsec:dmcl_cons}

While the Multi-View Query--Target Alignment objective anchors each query view to the target, it does not explicitly constrain the relationship between the textual query and its generated visual proxy. Due to the domain gap, the generated image \ignore{$I_n$}$I_{n,i}$ may deviate from the textual intent (e.g., hallucinating objects not mentioned in the text). We therefore introduce a text--diffusion consistency objective (indicated by orange arrows in Figure 3).

\myheader{Feature-level Consistency}
To explicitly couple the textual and diffusion query representation in the shared embedding space and reduce cross-view drift, we treat the reformulated text $T_{n,i}$ and its corresponding generated image $I_{n,i}$ as a positive pair and apply an InfoNCE loss:
\begin{equation}
\mathcal{L}_{\textsf{cons}} = \mathcal{L}_{\textsf{NCE}}(z_n^{\textsf{T}}, z_n^{\textsf{D}}; \tau_{\textsf{TD}}).
\end{equation}

\myheader{Distribution-level Agreement}
Let $p^{\textsf{T}}_{i,\cdot}$ and $p^{\textsf{D}}_{i,\cdot}$ denote the softmax retrieval distributions over batch targets induced by $z_{n,i}^{\textsf{T}}$ and $z_{n,i}^{\textsf{D}}$. We penalize their disagreement via Jensen--Shannon divergence:
\begin{equation}
    \mathcal{L}_{\textsf{dist}}
    = \frac{1}{N}\sum_{i=1}^N \mathrm{JS}\!\left(p^{\textsf{T}}_{i,\cdot}\;\|\;p^{\textsf{D}}_{i,\cdot}\right).
\end{equation}

\subsubsection{Overall Training Objective}
\label{subsubsec:dmcl_total}
The final training objective combines both objectives:

\begin{equation}
    \mathcal{L}_{\textsf{total}}
    =
    \mathcal{L}_{\textsf{align}}
    + \beta_{\textsf{cons}}\mathcal{L}_{\textsf{cons}}
    + \beta_{\textsf{dist}}\mathcal{L}_{\textsf{dist}}.
\end{equation}

\subsection{Why DMCL Enables Semantic Filtering}
\label{subsec:dmcl_analysis}

As argued in Section~\ref{subsec:why_filter}, contrastive alignment acts as a semantic filter: it consistently rewards the ``intent'' component shared across query views while suppressing the ``hallucination'' component that varies across query views. DMCL instantiates this principle in DAI-TIR via the two contrastive objectives described above.

\myheader{The Alignment Objective as task grounded filtering}
The Multi-View Query--Target Alignment objective anchors \emph{each} query view to the same target image. This is a direct realization of the alignment argument in \ric{Section~}\ref{subsec:why_filter}, but applied to three views simultaneously. Text--target alignment preserves explicit semantic constraints expressed in the dialogue. Diffusion--target alignment injects global structure and style cues when they are helpful, while discouraging diffusion hallucination that do not consistently increase agreement with the target. Fused--target alignment further aggregates complementary evidence and learns to emphasize signals that are jointly predictive. Hard negative mining strengthens the filtering effect by penalizing hallucination-sensitive directions that increase similarity to confusable candidates. Overall, this objective ensures that diffusion-induced signal that persists in the representation is \emph{retrieval-useful} and remains stable under discrimination pressure.

\myheader{The Consistency objective as semantic anchoring against cross view drift}
Query--target alignment alone does not guarantee that the diffusion view remains faithful to the dialogue semantics, because a hallucinated proxy might still correlate with the target within a batch. The Text--Diffusion Consistency objective addresses this by explicitly aligning the diffusion embedding to the text embedding. The feature level consistency loss reduces representation drift between the two query views, and the distribution level agreement further enforces that they induce similar retrieval score distributions over candidates. Together, these constraints implement the semantic anchoring intuition in \ric{Section}~\ref{subsec:why_filter}, pushing diffusion specific hallucinations that are not supported by text toward the null space, while preserving shared intent signals.

\myheader{Why two objectives are necessary}
The two objectives play distinct roles. The alignment objective provides task grounded supervision that preserves discriminative structure and prevents trivial collapse, while the consistency objective provides a clean semantic reference that reduces diffusion domain shift and hallucination induced conflicts. Their combination yields a robust embedding space where diffusion augments the representation without distorting the underlying retrieval intent.

\subsection{Diffusion-Augmented Training Data Construction}
\label{subsec:data}

To equip the model with the capability to effectively filter diffusion induced \noisename, we construct a Diffusion Augmented Interactive Dataset based on the training set of VisDial dataset~\cite{das2017visual}, named DA-VisDial. Specifically, as detailed in Algorithm~\ref{algo:data-construction}, we employ an LLM to reformulate the accumulated dialogue into a concise description with a few-shot in-context learning prompting strategy~\cite{liu2022few, ma2023fairness, yao2024more}, which then conditions a diffusion model to synthesize the proxy image. The resulting dataset $ \mathcal{D}_{\mathrm{triple}} = \{ (C_{n,i}, I_{n,i}, I_i) \mid 1 \le n \le N\ , \mid 1 \le i \le M \} $,  containing processed samples across all rounds, will be publicly released to support further research.

\begin{algorithm}
    \caption{Diffusion-Augmented Interactive  Text-to-Image Retrieval Dataset Construction}
    \label{algo:data-construction}
    \small
    \linespread{0.95}\selectfont
    \SetNlSty{textbf}{\normalsize}{}
    \DontPrintSemicolon 
    \SetAlgoNoLine 
    \SetKwInput{KwNotation}{Notation}  
    \textbf{Notation:}\\
    \Indp 
     $\mathcal{R}(\cdot)$ : LLM for dialogue reformulation\\
     $\mathcal{G}(\cdot)$ : text-to-image diffusion generator\\
     $T$ : maximum dialogue rounds\\
     $M$ : number of instances in datasets\\
    \Indm 
    Initialize diffusion-augmented datasets $\mathcal{D}_{\mathrm{triple}} \leftarrow \emptyset$\;
    \For{$i = 1$ \KwTo $M$}{
        Initialize dialogue context $ C_{0,i}  =  D_{0,i} $\\
        \For{$n = 1$ \KwTo $N$}{
            System Inquiry: Question $Q_{n,i}$ \\
            User Response: Corresponding Answer $ A_{n,i} $ \\
            Update dialogue context: $ C_{n,i} = C_{n-1,i} \cup \{Q_{n,i} , A_{n,i}\} $\; 
            Context Reformulation: $S_{n,i} = \mathcal{R}(C_{n,i}) $\;
            Generated Reference Image: $I_{n,i} = \mathcal{G}(S_{n,i})$\; 
            Update datasets: $\mathcal{D}_{\mathrm{triple}} = \mathcal{D}_{\mathrm{triple}} \cup \{ (C_{n,i}, I_{n,i}, I_i) \}$\;
        }
    }
    \KwRet{$\mathcal{D}_{\mathrm{triple}} = \{ (C_{n,i}, I_{n,i}, I_i) \mid 1 \le n \le N\ , \mid 1 \le i \le M \} $}
\end{algorithm}

\section{Experimental Setup}
\myheader{Datasets and Training Details} 
We conduct all training on our DA-VisDial mentioned in Section\ref{subsec:data}.
Regarding implementation specifics, we instantiate the Reformulator with BLIP-3~\cite{xue2024xgen} and employ Stable Diffusion 3.5~\cite{esser2024scaling} as the diffusion generator. We adopt the BEiT-3 base model~\cite{wang2023image} as the multimodal encoder and initialize it with publicly available pretrained weights. 


We report results on the VisDial~\cite{das2017visual} validation split and on extra four I-TIR datasets constructed by ChatIR~\cite{levy2023chatting}: \textit{ChatGPT\_BLIP2}, \textit{HUMAN\_BLIP2}, \textit{Flan-Alpaca-XXL\_BLIP2} and \textit{PlugIR\_dataset}, as zero-shot testing cases. 
Following previous works~\cite{levy2023chatting,lee2024interactive, long2025diffusion},  we adopt the accumulated recall at rank 10 (Hits@10) as our primary evaluation metric, i.e., the proportion of test queries for which the target image has appeared at least once within the top-10 retrieved images up to a given dialogue round. The Code for all experiments and the training dataset are available at:
\ric{\begin{center}
        \underline{\emph{\url{https://github.com/longkukuhi/DMCL}}}
\end{center}}




\myheader{Baselines and DMCL} We compare our DMCL framework with three baselines, namely ZS, ChatIR and PlugIR-CR. We evaluate each backbone under two inference settings: standard text-only retrieval and diffusion-augmented retrieval (denoted by the suffix \_DAR). For all \_DAR variants, we adopt a consistent weighted fusion strategy as in DAR~\cite{long2025diffusion}.
\begin{itemize}
    \item \textbf{BEiT-3/BEiT-3\_DAR:} The BEiT-3 baseline directly uses the pre-trained BEiT-3 weights~\cite{wang2023image} without any additional training on I-TIR datasets, serving as a zero-shot reference. The same backbone also initializes our DMCL model, so this baseline provides a direct ablation that isolates the impact of the DMCL training framework beyond the encoder architecture.

    \item \textbf{ChatIR/ChatIR\_DAR:} ChatIR~\cite{levy2023chatting} fine-tunes a text encoder on VisDial for I-TIR and serves as a strong I-TIR task trained baseline that relies on standard text queries. Building on this backbone, ChatIR\_DAR applies the same diffusion augmented fusion strategy described above to incorporate generated visual proxies.

    \item \textbf{PlugIR} To reproduce PlugIR~\cite{lee2024interactive} while ensuring a fair comparison in model scale, we rewrite each multi-round dialogue across the four benchmarks into a single caption style query, and then perform retrieval using a BLIP base model~\cite{li2022blip} fine-tuned on MSCOCO~\cite{lin2014microsoft}.

\end{itemize}

\begin{figure*}[t]
    \centering
    \setlength{\abovecaptionskip}{3pt} 
    \setlength{\belowcaptionskip}{0pt}

    \begin{subfigure}[b]{0.32\linewidth}
        \centering
        \captionsetup[subfigure]{singlelinecheck=false, justification=raggedright}
        
        \includegraphics[width=\linewidth, trim=5pt 3pt 5pt 0, clip]{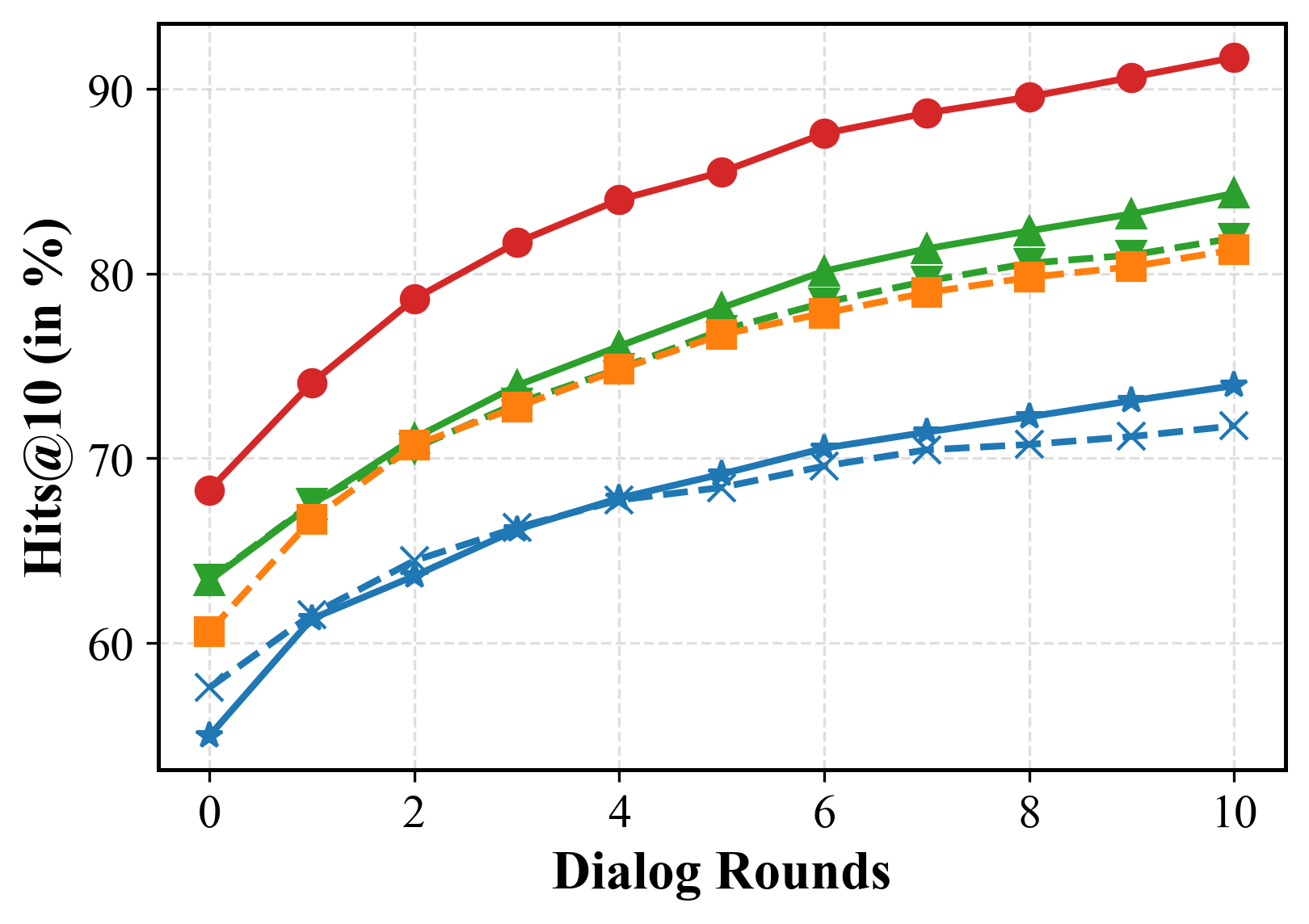}
        \vspace{-0.6cm}
        \captionsetup{margin={-1.6em,0pt}}
        \caption{VisDial}
        \label{fig:visdial}
    \end{subfigure}
    \hfill
    \begin{subfigure}[b]{0.32\linewidth}
        \centering
        \includegraphics[width=\linewidth, trim=5pt 3pt 5pt 0, clip]{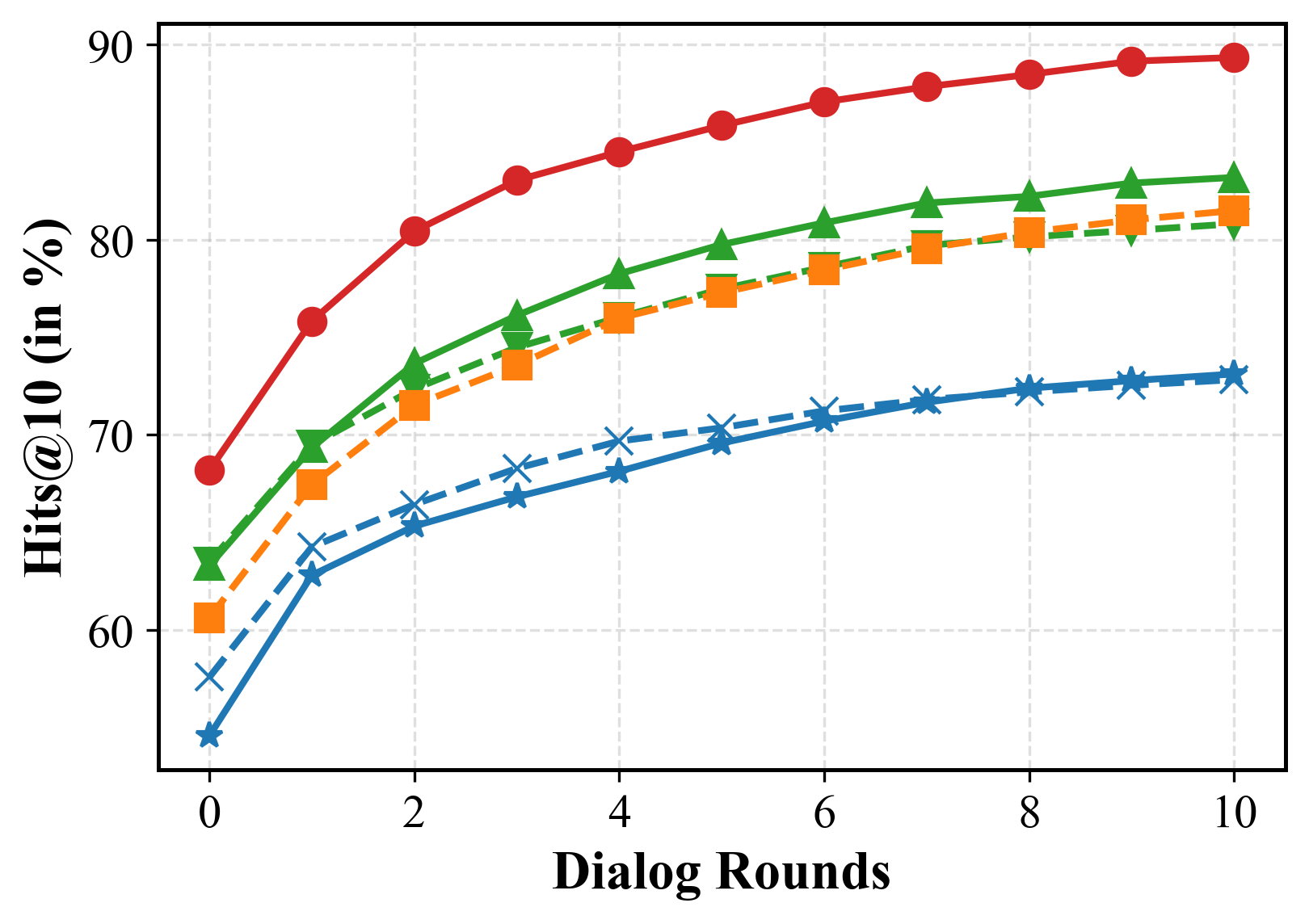}
        \vspace{-0.6cm}
        \captionsetup{margin={-1.0em,0pt}}
        \caption{ChatGPT\_BLIP2}
        \label{fig:chatgpt}
    \end{subfigure}
    \hfill
    \begin{subfigure}[b]{0.32\linewidth}
        \centering
        \includegraphics[width=\linewidth, trim=5pt 3pt 5pt 0, clip]{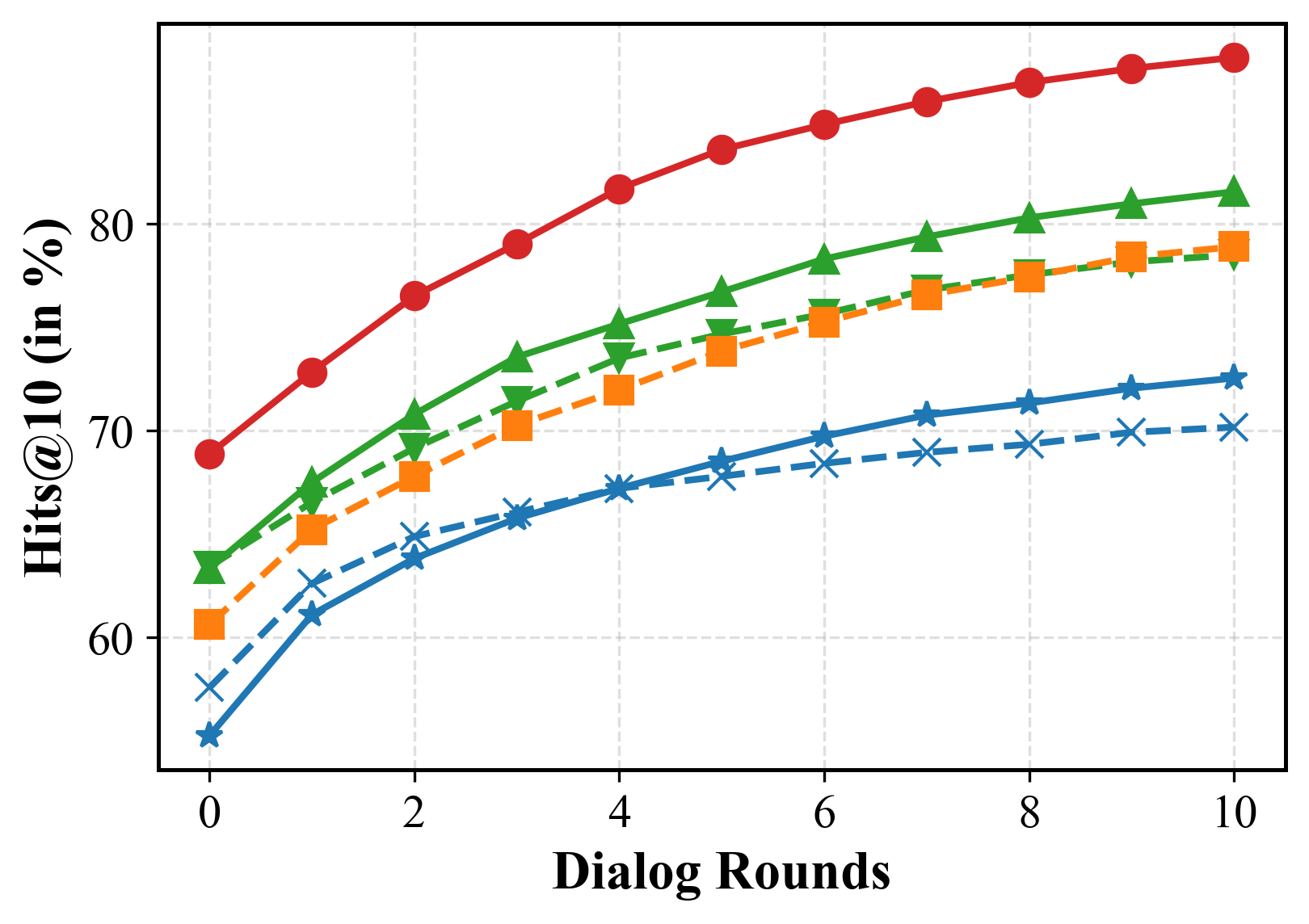}
        \vspace{-0.6cm}
        \captionsetup{margin={-1.8em,0pt}}
        \caption{Human\_BLIP2}
        \label{fig:human}
    \end{subfigure}
    
    \vspace{-0.1cm} 
    \begin{subfigure}[b]{0.32\linewidth}
        \centering
        \includegraphics[width=\linewidth, trim=5pt 3pt 5pt 0, clip]{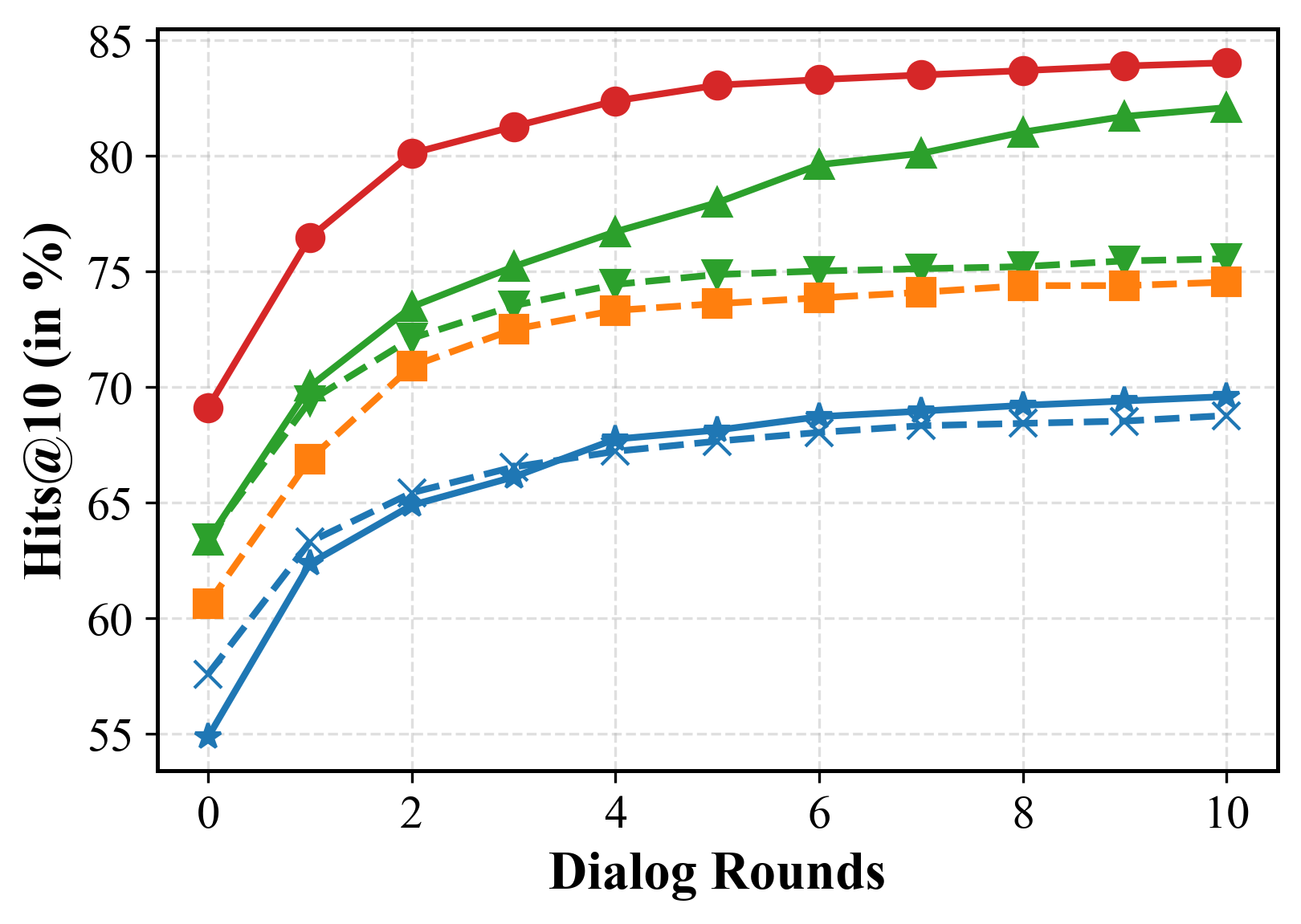}
        \vspace{-0.6cm}
        \captionsetup{margin={-1.5em,0pt}}
        \caption{Flan-Alpaca-XXL\_BLIP2}
        \label{fig:flan}
    \end{subfigure}
    \hfill
    \begin{subfigure}[b]{0.32\linewidth}
        \centering
        \includegraphics[width=\linewidth, trim=5pt 3pt 5pt 0, clip]{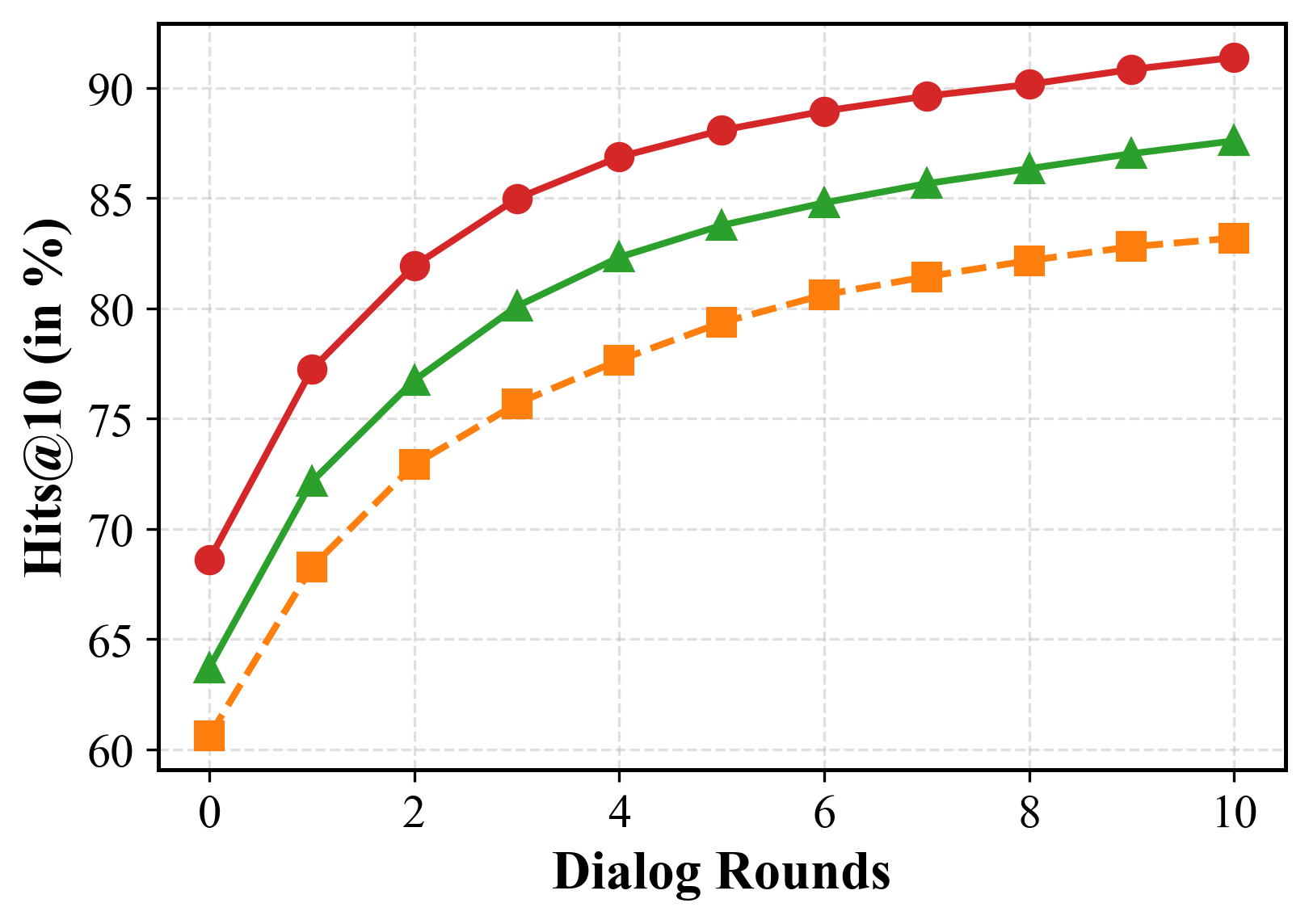}
        \vspace{-0.6cm}
        \captionsetup{margin={-1.2em,0pt}}
        \caption{PlugIR\_Dataset}
        \label{fig:plugir}
    \end{subfigure}
    \hfill
    \begin{subfigure}[b]{0.30\linewidth}
        \centering
        \raisebox{2mm}
        {\includegraphics[width=0.95\linewidth, trim=20pt 30pt 60pt 20pt, clip]{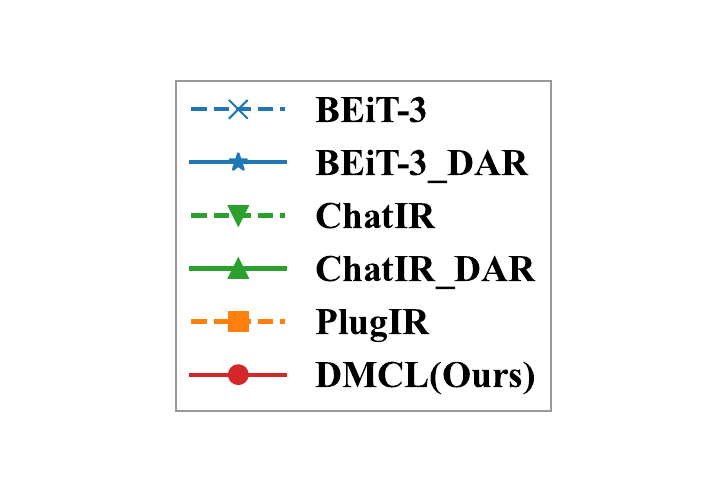}}
        \vspace{6.5pt}
        \caption*{\phantom{Placeholder}}
    \end{subfigure}
    \vspace{-0cm} 
    \caption{The overall interactive retrieval performance (measured by cumulative Hits@10) across five benchmarks. Our proposed DMCL consistently outperforms baselines across all dialogue rounds.}
    \label{fig:overall_performance}
    \vspace{-3mm} 
\end{figure*}

\section{Experimental Results}
\label{sec:experiments}

\vspace{5pt}
\subsection{In-Distribution Performance Comparison}
\vspace{3pt}
We first evaluate retrieval performance on the VisDial validation split to assess the training effect of DMCL under an in-distribution setting, ensuring a fair comparison with baselines trained for I-TIR on the same dataset. Since the training and validation data come from the same source, we refer to this as \emph{in-distribution} evaluation.

As shown in Figure~\ref{fig:visdial}, the diffusion augmented baseline (BEiT-3\_DAR) underperforms the text only BEiT-3 model in early dialogue rounds, supporting our observation that diffusion generated visual proxies can introduce \noisename and cross modal inconsistencies that harm retrieval. In contrast, DMCL achieves the best performance across all rounds. Its advantage appears from the beginning of the interaction and widens as the dialogue progresses. Compared with the previous state of the art (ChatIR\_DAR), DMCL improves by 4.90\% at round 0 and 7.37\% at round 10. Overall, these results suggest that DMCL effectively mitigates diffusion induced \noisename while leveraging multi round feedback more reliably, yielding sustained gains in retrieval performance.



\subsection{Out-of-Distribution Generalization}
\vspace{3pt}
To assess whether DMCL generalizes beyond the VisDial training distribution, we conduct zero-shot evaluation on four out-of-distribution I-TIR benchmarks: ChatGPT\_BLIP2, HUMAN\_BLIP2, Flan-Alpaca-XXL\_BLIP2, and PlugIR\_dataset. These benchmarks differ from VisDial in dialogue style, linguistic redundancy, and the expression of visual references, thereby providing a diverse testbed for distribution shift. 

As shown in Figure~\ref{fig:chatgpt}--\ref{fig:flan}, DMCL consistently achieves the best cumulative Hits@10 across all dialogue rounds on each benchmark. Specifically, on ChatGPT\_BLIP2 and HUMAN\_BLIP2, DMCL surpasses the strongest diffusion-augmented baseline (ChatIR\_DAR~\cite{long2025diffusion}) by a substantial margin, achieving a Hits@10 improvement of 6.15\% and 6.49\% at the final round, respectively. These results quantitatively verify DMCL’s strong robustness to dialogue-style variation and generative perturbations.

We further analyze generalization on the PlugIR benchmark, which differs from the above datasets in that it is constructed by an interactive pipeline. Specifically, PlugIR~\cite{lee2024interactive} introduces a context-aware dialogue generation module (CDG) that leverages current retrieval candidates to ask more discriminative, non-redundant questions, together with a context reformulation module (CR) that rewrites multi-round dialogue histories into caption-style queries optimized for vision--language retrievers. Since CDG requires access to a live retrieval loop and cannot be applied to the fixed dialogues of the standard benchmarks, we perform comparisons at two levels.

\vspace{2mm}\noindent \textbf{PlugIR-CR on fixed-dialogue benchmarks.}
Following PlugIR, we apply the CR module to the four standard benchmarks \ric{used previously}, rewriting each dialogue round into a caption-style query and performing retrieval with a BLIP base model~\cite{li2022blip} fine-tuned on MSCOCO, whose parameter count is comparable to ours. As reported in Figure~\ref{fig:visdial}--Figure~\ref{fig:flan}, PlugIR-CR performs competitively with strong text-only baselines (e.g., ChatIR), but generally underperforms approaches that additionally leverage visual signals (e.g., ChatIR\_DAR). \ric{Meanwhile, we observe that} DMCL consistently delivers the best performance across datasets and dialogue rounds.

\vspace{2mm}\noindent \textbf{PlugIR full pipeline on PlugIR\_dataset.}
We then evaluate on the PlugIR\_dataset, where queries are generated by the full PlugIR pipeline (CDG+CR). As shown in Figure~\ref{fig:plugir}, DMCL remains the top-performing method even under these carefully optimized summary-style queries, reaching 91.38\% Hits@10 at round 10 and surpassing both ChatIR\_DAR and the PlugIR pipeline by +3.78\% and +8.19\% points respectively. This suggests that DMCL is not tied to a particular query form; instead, it effectively captures complementary latent semantics while suppressing distracting or spurious content, leading to consistent retrieval gains.   
Overall, the improvements remain clear and stable, demonstrating that DMCL generalizes well across diverse dialogue structures and query formulations.

\subsection{Effect of two Multi-View training Objectives}
We also conducted a small ablation study on DMCL to evaluate the impact of the two proposed objectives that we integrate into the loss function. We observed that the multi-view query–target alignment objective (see Section~\ref{subsubsec:dmcl_align}) provides the majority of the gain by anchoring each query view to the ground-truth target, specifically, this objective alone yields a significant improvement of 6.45\% in cumulative Hits@10 compared to the ChatIR\_DAR baseline. Adding the text–diffusion consistency objective (see Section~\ref{subsubsec:dmcl_cons}) further reduces cross-view drift between the textual query and its diffusion proxy image, yielding an additional 1.2\% improvement in cumulative Hits@10 at round 10.

\begin{figure*}[t] 
    \centering
    \includegraphics[width=1.0\textwidth]{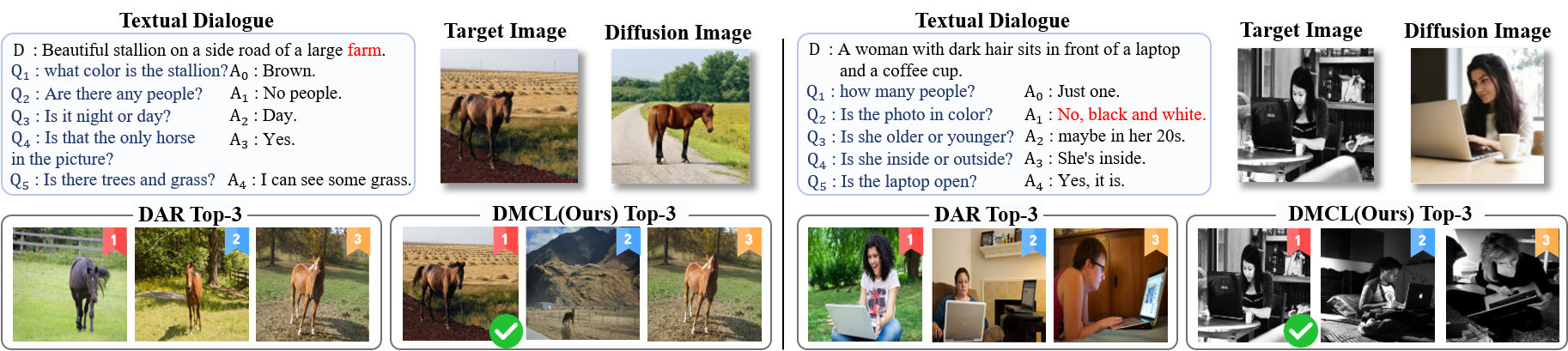}
    \vspace{-6mm}
    \caption{Qualitative Evidence for Conflict Detail Filtering: DAR vs. DMCL(Ours)}
    \vspace{-3mm}
    \label{fig:Qualitative} 
\end{figure*}

\begin{figure} 
    \vspace{-2mm}
    \centering
    \includegraphics[width=\linewidth]{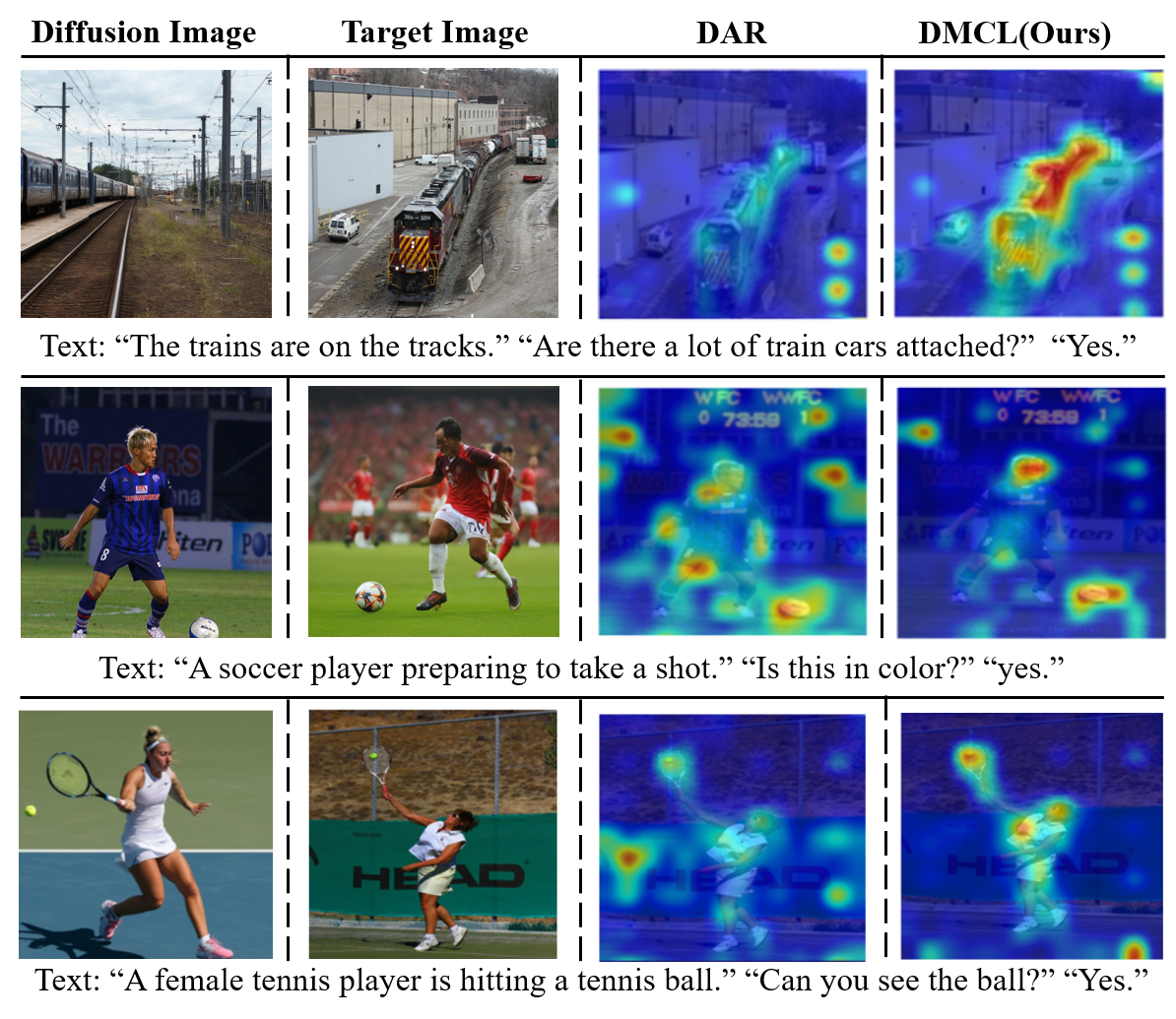}
    \vspace{-8mm}
    \caption{\textbf{Visualization of attention relevancy maps.}}
    \vspace{-4mm}
    \label{fig:attention_visualization} 
\end{figure}

\section{DMCL Qualitative Analysis}
\subsection{Ignoring Diffusion Hallucinated Cues}
Beyond the quantitative results in Section~\ref{sec:experiments}, we provide qualitative analysis to highlight DMCL’s improved fine-grained alignment and retrieval discrimination in Figure~\ref{fig:Qualitative}. Diffusion-generated images may introduce dialogue-inconsistent appearance cues, which bias the retrieval models toward similar-looking candidates that deviate from the target intent. In contrast, DMCL denoises conflicting cues during text–diffusion fusion and more reliably promotes intent-consistent results.
In left case in Figure~\ref{fig:Qualitative}, the diffusion generated image exhibits a green background that distracts the baseline, while DMCL suppresses this irrelevant background feature and improves the overall consistency of top-ranked candidates with the dialogue constraints.
In right case in Figure~\ref{fig:Qualitative}, the dialogue specifies a black-and-white indoor photo with a woman and a laptop, whereas the diffusion generated image is in color, leading the baseline to favor color scenes. DMCL captures the intended semantics and mitigates the influence of diffusion-induced spurious cues, resulting in top-ranked items that are collectively better aligned with the query description. This further supports our framework's ability to denoise multi-view signals and extract the correct intent cues for more reliable retrieval gains.


\vspace{-4mm}
\subsection{Improved Model Focus on Subjects}
To further mechanistically validate DMCL's capability to filter diffusion-induced \noisename and accurately extract user intent, we adopt the transformer explainability method of Chefer et al.~\cite{chefer2021transformer,chefer2021generic} to generate attention heatmaps showing which visual regions the model primarily attends to when computing the similarity score between the fused query and the ground-truth target image.

As illustrated in Figure~\ref{fig:attention_visualization}, the DAR Baseline exhibits a scattered attention distribution, often failing to ground the semantic subject specified in the dialogue. This issue is exacerbated when the diffusion generated image introduces visual discrepancies or inconsistent cues with the target: on the one hand, attention tends to spread to task-irrelevant background regions (e.g., in first and second line, the DAR method responds to multiple semantically unrelated areas); on the other hand, conflicting cues from the generated image may even induce erroneous focus (e.g., in third line, the baseline places excessive attention on an incorrect tennis-ball location in the diffusion generated image). In contrast, across various scenarios involving static objects or action descriptions, DMCL appears to effectively filter out conflicting and irrelevant information from the generated images, and thereby more consistently focus on the subject and critical regions.

\begin{figure}[t]
    \centering
    \includegraphics[width=\linewidth]{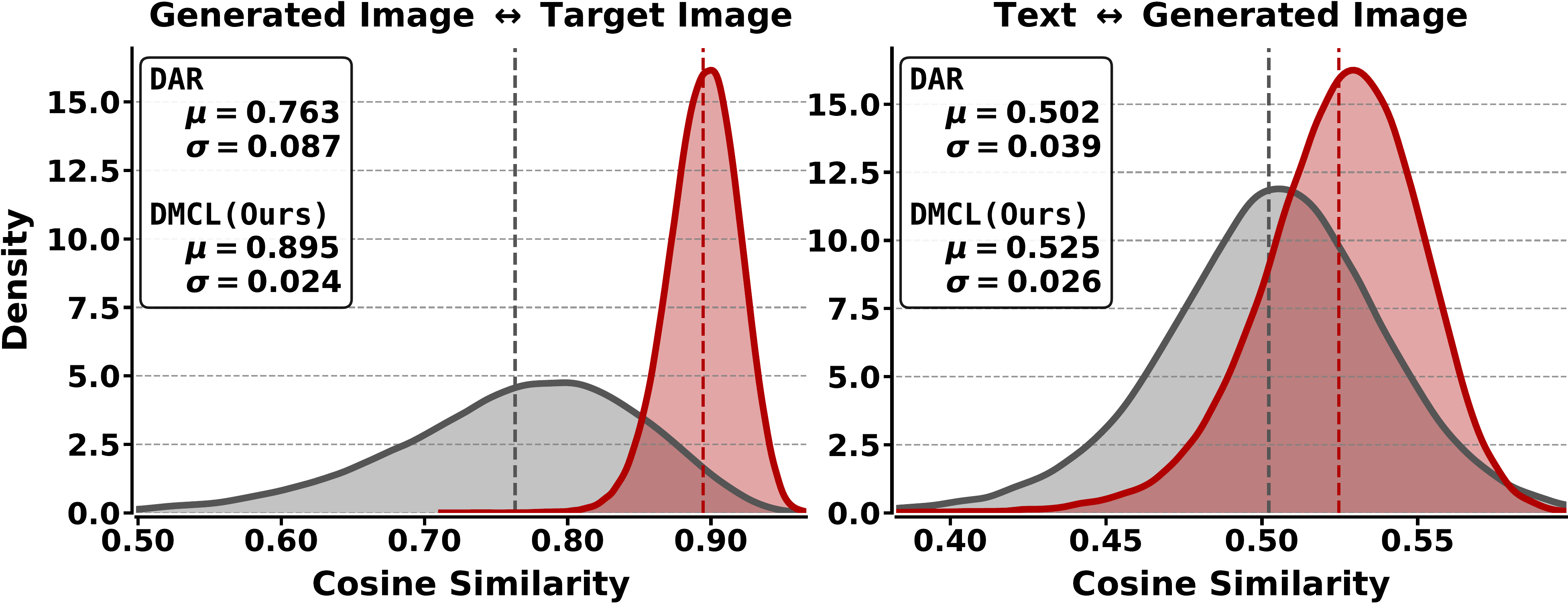}
    \vspace{-0cm} 
    \includegraphics[width=0.5\linewidth]{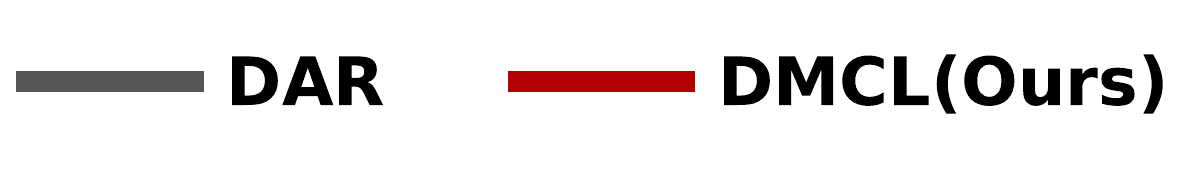}
    \vspace{-5mm} 
    \caption{Comparison of cross-modal semantic alignment capabilities between DAR and our DMCL framework.}
    \label{fig:sim_comparison}
    \vspace{-5mm} 
\end{figure}

\subsection{Better Semantic Alignment}
We further probe the geometry of the embedding space to clarify how the model mitigates diffusion \noisename while preserving the underlying intent. Specifically, we visualize the density of cosine similarities for positive pairs, examining both (i) \ric{\emph{generated image} to \emph{target image}} semantic consistency and (ii) \ric{\emph{text} to \emph{generated image}}  semantic consistency. As shown in Figure~\ref{fig:sim_comparison}, the DAR baseline exhibits a broad, high-variance similarity distribution with overall lower scores and noticeable instability, most prominently in the visual--visual alignment between the generated and target images. This dispersion indicates that DAR is more susceptible to \noisename, which in turn degrades retrieval performance. In contrast, DMCL produces a substantially tighter distribution that consistently shifts toward higher similarity values. This behavior suggests that the DMCL encoder acts as a semantic filter by attenuating generation-induced \noisename and amplifying task-relevant semantic cues, yielding more stable and faithful alignment with the target semantics.


\subsection{Limitations}
In this work, DMCL focuses on filtering \noisename to learn robust encoder representations, while query fusion still adopts a simple additive fusion scheme. Future work will explore more effective query fusion on multiple views to further improve retrieval performance

\section{Conclusion}

In this work, we propose Diffusion-aware Multi-view Contrastive Learning (DMCL), a hallucination-resilient training framework for Diffusion-Augmented Interactive Text-to-Image Retrieval (DAI-TIR). Through theoretical analysis and extensive empirical evidence, we show that DMCL acts as a semantic filter: it suppresses diffusion-induced \noisename while preserving and amplifying intent-relevant signals for retrieval. To support future research, we release a publicly available large-scale DAI-TIR training dataset. Comprehensive experiments across multiple settings validate the effectiveness and robustness of DMCL, establishing it as a general framework that better reconciles generative augmentation with discriminative retrieval.

\bibliographystyle{plain}
\bibliography{paper}

\end{document}